\documentclass[twocolumn,showpacs,preprintnumbers,amsmath,amssymb]{revtex4}

\usepackage{graphicx}
\usepackage{dcolumn}
\usepackage{bm}

\usepackage{epsfig}

\def\be{\begin{equation}}
\def\ee{\end{equation}}

\newcommand{\ba}{\begin{eqnarray}}  
\newcommand{\bad}{\begin{array}{ccc}}
\newcommand{\bea}{\begin{equation} \begin{array}{c}}
\newcommand{\eea}{ \end{array} \end{equation}}
\newcommand{\ea}{\end{eqnarray}}

\newcommand{\Ord}{\ensuremath{{\cal O}}}

\newcommand{\eL}{\ensuremath{{\bf e}}_{\mathrm L}}

\newcommand{\eN}{\ensuremath{{\bf e}}_{\mathrm N}}
\newcommand{\eT}{\ensuremath{{\bf e}}_{\mathrm T}}
\newcommand{\ei}{\ensuremath{{\bf e}}_{\mathrm i}}

\newcommand{\pq}{\ensuremath{{\bf p}}_{q}}
\newcommand{\plep}{\ensuremath{{\bf p}}_{l^-}}
\newcommand{\vecn}{\ensuremath{{\bf n}}}
\newcommand{\PL}{\ensuremath{{P}_{\rm L}}}
\newcommand{\PN}{\ensuremath{{P}_{\rm N}}}
\newcommand{\PT}{\ensuremath{{P}_{\rm T}}}
\newcommand{\PI}{\ensuremath{{P}_{\rm i}}}
\newcommand{\ACP}{\ensuremath{{A}_{\rm CP}}}
\newcommand{\ACPL}{\ensuremath{\delta{A}_{\rm CP}^{\rm L}}}
\newcommand{\ACPN}{\ensuremath{\delta{A}_{\rm CP}^{\rm N}}}
\newcommand{\ACPT}{\ensuremath{\delta{A}_{\rm CP}^{\rm T}}}
\newcommand{\lu}{\ensuremath{{\lambda}_{u}}}
\newcommand{\BR}{\ensuremath{\operatorname{BR}}}

\newcommand{\shat}{\ensuremath{{\hat s}}}
\newcommand{\mchat}{\ensuremath{{\hat m}_c}}
\newcommand{\muhat}{\ensuremath{{\hat m}_u}}

\newcommand{\mqhat}{\ensuremath{{\hat m}_q}}
\newcommand{\mlhat}{\ensuremath{{\hat m}_l}}

\newcommand{\gq}{\ensuremath{g(\mqhat,\shat)}}

\newcommand{\Imag}{\ensuremath{{\operatorname{Im}}}}
\newcommand{\Real}{\ensuremath{{\operatorname{Re}}}}
\newcommand{\BABAR}{\ensuremath{\text{BaBar}\ }}
\newcommand{\BELLE}{\ensuremath{\text{Belle}\ }}
\newcommand{\Cixeff}{\ensuremath{C_9^{\rm eff}}}
\newcommand{\Cviieff}{\ensuremath{C_7^{\rm eff}}}

\newcommand{\lsim}{{\;\raise0.3ex\hbox{$<$\kern-0.75em\raise-1.1ex
\hbox{$\sim$}} \;}} 
\newcommand{\gsim}{{\;\raise0.3ex\hbox{$>$\kern-0.75em\raise-1.1ex
\hbox{$\sim$}} \;}} 

\begin{document}


\title{$CP$ violation in polarized $b \to d l^+ l^-$: A Detailed 
Standard Model Analysis}

\author{K.\ S.\ Babu}
\email{babu@hep.phy.okstate.edu}
\affiliation{Department of Physics, Oklahoma State University, 
Stillwater, Oklahoma 74078, USA}

\author{K.\ R.\ S.\ Balaji}
\email{balaji@hep.physics.mcgill.ca}
\affiliation{Department of Physics, McGill University, 
Montr{\'e}al, Qu{\'e}bec, H3A 2T8, Canada}

\author{\underline{I.\ Schienbein}}
\email{schien@mail.desy.de}
\affiliation{DESY, Notkestrasse 85, 22603 Hamburg, Germany}

\date{\today}

\begin{abstract}
The electroweak $CP$ violating rate asymmetries 
in the decay $b \to d l^+ l^-$ are reexamined in detail and updated.
In particular, the rate asymmetries are studied when one of the final
state leptons is polarised. We find an estimate for the asymmetry of 
$(5\div 15)\%$ in the polarised decay spectrum which is close to known
results for the unpolarised case. Interestingly, in the region separating the
$\rho-\omega$ and $c \bar c$ resonances, which is also theoretically cleanest,
the polarised contribution
to the asymmetry is larger than the unpolarised result. 
A $3\sigma$ signal for direct CP violation, requires about $10^{10}$ $B\bar B$ 
pairs at a B factory. In general these results
indicate an asymmetric contribution from the individual polarisation states
to the unpolarised $CP$ asymmetry; an atribute for any new physics searches.
\end{abstract}

\maketitle

\section{\label{sec:intro} Introduction}
In the standard model (SM), the rare decays which involve
flavor changing neutral current decays are important probes 
for new physics \cite{Buras:2001pn}. In general, due to the well
known GIM mechanism \cite{Glashow:1970gm} these processes are strongly
suppressed along with the usual CKM suppression. All these features
therefore point to potential sources or act as rich test ground for any 
new physics, such that any deviation from the SM expectations is an 
unambiguous signature for new physics. 
One of the key observables involving FCNC is the radiative B decay,
$B \rightarrow X_s \gamma$.
Measurements of the branching ratio for this decay 
\cite{Battaglia:2003in}
are in very good agreement with the SM calculations 
\cite{Gambino:2001ew,Buras:2002tp} 
at the current level of accuracy
imposing strong constraints on new physics scenarios.

The parton level rare semi-leptonic decays, 
$b \rightarrow q l^+l^-$ ($q=d,s$), can 
provide alternative sources to discover new physics 
where
in particular the lepton pair gives easy access to 
decay spectra in dependence of the invariant mass of the lepton pair
providing detailed dynamical information.
Moreover, appropriate experimental cuts allow to separate out
theoretically clean regions of the phase space.
The semi-leptonic decays 
are described by QCD corrected effective Hamiltonians 
which are expanded in a set of effective operators multiplied by 
so called Wilson coefficients acting as couplings and
being perturbatively calculable.
In new physics scenarios the Wilson couplings can get modified
and, in addition, new operator structures can arise.
In such decays, the standard  
observables like the decay rate, lepton polarization 
asymmetries and the forward-backward asymmetry 
depend on different quadratic combinations of the 
Wilson coefficients 
and can be studied kinematically as a function
of the invariant di-lepton mass.  
Therefore, detailed measurements of these observables
provide extensive tests of the effective Hamiltonian
and hence of the SM and new physics scenarios.
Before the B factories such as \BABAR and \BELLE 
the best experimental limits 
for the inclusive branching ratios $BR(b \to s l^+ l^-)$ 
with $l = e,\mu$ as measured by CLEO \cite{Glenn:1998gh} 
have been an order higher than the SM estimates \cite{Ali:1997bm}. 
However, the first measurement of this decay has been reported 
by \BELLE \cite{Kaneko:2002mr} and is in agreement with the SM expectations
and hence further constrains any extensions to the SM. 

In addition to the above observables it is possible
to construct $CP$ violating rate asymmetries which
are sensitive to further different combinations of 
Wilson couplings.
Needless to say, that apart from new physics searches,
$CP$ violating effects in rare $b$-decays 
are also interesting in its own right.
In these proceedings we summarize the main results of our 
recent analysis of $CP$ violating effects in the decay 
$b \to d l^+ l^-$ within the SM, including the case when 
one of the leptons is polarized \cite{babu:2003ir}.

Standard model $CP$ violation in the decays $b \to q l^+ l^-$ has 
been studied previously in Refs.\ \cite{Kruger:1997dt,Ali:1998sf}
for unpolarized leptons. 
Very recently, 
a study of $CP$ violation
in the polarized decay $b\to d l^+ l^-$ has also been performed
in a model independent framework \cite{Aliev:2003hw}.

\section{Theoretical Rate Estimates}\label{sec:framework}

The parton level process with the QCD corrected effective Hamiltonian 
describing the decay $b \to d l^+ l^-$  can be described by the matrix 
element \cite{Kruger:1997dt,Buras:1995dj}
\begin{eqnarray}
M &=&  
K
\Big[{\Cixeff}(\bar{d} \gamma_\mu P_{L} b) \bar{l} \gamma^{\mu} l  
    +C_{10}(\bar{d} \gamma_\mu P_{L} b)  \bar{l} \gamma^{\mu} \gamma^{5}l 
\nonumber\\*
& & 
\phantom{K}
- 2 {\Cviieff} \bar{d} i\sigma_{\mu\nu} \frac{q^{\nu}}{q^2} (m_{b}P_{R} + 
  m_{d}P_{L})b \bar{l} \gamma^{\mu} l\Big],
\label{eq:heff}
\end{eqnarray}
with $K := \frac{G_{F} \alpha V_{tb}V_{td}^*}{\sqrt{2} \pi}$.
In Eq.\ \eqref{eq:heff}
the notations employed are the standard ones and $q$ denotes the 
four momentum of the lepton pair.

In the SM, except for $\Cixeff$, the Wilson couplings are real and analytic 
expressions can be found in the literature 
\cite{Buras:1995dj,Buchalla:1996vs,misiak}. 
In our analysis \cite{babu:2003ir} we use 
\be
\begin{gathered}
\Cviieff = - 0.310\ ,\quad C_{10} = -4.181\ .
\label{eq:coeff}
\end{gathered}
\ee
The effective coefficient $\Cixeff$ 
can be parametrized in the following way \cite{Kruger:1997dt}:
\be
\Cixeff = \xi_1 + \lu \xi_2\ , \quad \lu = 
\frac{V_{ub}V_{ud}^*}{V_{tb}V_{td}^*}~,
\label{eq:c9eff}
\ee
with 
\begin{eqnarray}
\xi_1 &\simeq& 4.128 + 0.138\ \omega(\shat) 
+ 0.36\ g(\mchat, \shat),
\\
\xi_2 &\simeq& 0.36\ [g(\mchat, \shat) - g(\muhat, \shat)]~.
\label{approx}
\end{eqnarray}
Here, $\shat = q^2/m_b^2$ and $\mqhat = m_q/m_b$ are dimensionless variables 
scaled with respect to the bottom quark mass. 
The function $\omega(\shat)$ represents one loop corrections
to the operator $O_9$ \cite{jezabek:1989ja} and the function $g(\mqhat, \shat)$
represents the corrections to the four-quark 
operators $O_1-O_6$ \cite{misiak}.

In addition to the short distance contributions described above,
the decays $B \to X_{d} l^{+} l^{-}$
also receive long distance
contributions from the tree-level diagrams involving 
$u \bar u$, $d \bar d$, and $c \bar c$
bound states, 
$B \to X_d  (\rho, \omega, J/ \psi, \psi^{\prime}, ...) \to  X_d l^+l^-$. 
In our analysis the $\rho$, $\omega$, and $c\bar c$ resonances
have been taken into account by an appropriate modification
of the functions $\gq$.

\subsection{Differential decay spectrum: Unpolarized case}
Neglecting any low energy QCD
corrections ($\sim 1/m_b^2$) \cite{Falk:1994dh,Ali:1997bm} and 
setting the down quark
mass to zero, the unpolarized differential decay width as a function of the
invariant mass of the lepton pair is given by
\begin{equation}
\frac{d\Gamma}{d{\shat}} =  \frac{G_{F}^2 m_{b}^5 \alpha^2}{768
\pi^5}|V_{tb}V_{td}^{*}|^2 (1-\shat)\
\sqrt{1 - \frac{4 {\mlhat}^2}{\shat}}
\Delta(\shat)\, ,
\label{diffspec}
\end{equation}  
with the kinematic factors  
\begin{eqnarray}
\Delta(\shat) &=& \big[12\Real({\Cviieff} {\Cixeff}^*)F_1(\shat) +
\frac{4}{\shat}|{\Cviieff}|^2
F_2(\shat)\big] 
\nonumber\\
&&
\times
(1+\frac{2 \mlhat^2}{\shat})
+(|{\Cixeff}|^2 +|C_{10}|^2)F_3(\shat) 
\nonumber\\
&&+6\mlhat^2(|{\Cixeff}|^2 -|C_{10}|^2)F_1(\shat)~,
\nonumber\\ 
F_1(\shat) &=&  1 -\shat~,\nonumber\\
F_2(\shat) &=&  2 - \shat - \shat^2~,\nonumber\\ 
F_3(\shat) &=&  1
+\shat -2\shat^2  + (1-\shat)^2 \frac{2\mlhat^2}{\shat}~.
\label{eq:defFunpoll}
\end{eqnarray}
In Eq.(\ref{eq:defFunpoll}) the hat denotes all parameters
scaled with respect to the $b$ quark mass $m_b$. 
Note that the physical range for $\shat$ is 
given by $4\mlhat^2 \le \shat \le 1$.

As usual we remove uncertainties in Eq.\ \eqref{diffspec} due to the
bottom quark mass (a factor of $m_b^5$) by introducing the charged current 
semi-leptonic decay rate
\be
\Gamma(B \to X_c e \bar \nu_e)= \frac{G_{F}^2 m_{b}^5}{192 \pi^3}
|V_{cb}|^2 f(\mchat) \kappa(\mchat)
\ee
where $f(\mchat)$ and $\kappa(\mchat)$ represent the phase space
and the one-loop QCD corrections to the semi-leptonic decay and
can be found in \cite{Kruger:1997dt}.
Therefore the differential branching ratio can be written as
\ba
\frac{d \BR}{d \shat} =  \frac{\alpha^2}{4 \pi^2}
\frac{|V_{tb}V_{td}^{*}|^2}{|V_{cb}|^2} 
\frac{\BR(B \to X_c e \bar \nu_e)}{f(\mchat)\kappa(\mchat)}
 (1-\shat)\ a\ \Delta(\shat),\
\label{eq:BR}
\ea
with the threshold factor $a \equiv \sqrt{1 - \frac{4 {\mlhat}^2}{\shat}}$.

\subsection{Differential decay spectrum: Polarized case}
The potential richness of measuring the lepton polarisation was first 
realised by
Hewett \cite{Hewett:1996dk} and Kr\"uger and Sehgal \cite{Kruger:1996cv}.
These authors showed that additional independent information 
can be obtained on the quadratic functions of the effective Wilson couplings,
${\Cviieff}$, $C_{10}$ and ${\Cixeff}$. Defining a reference
frame with three orthogonal unit vectors $\eL$, $\eN$ and $\eT$, such that
\begin{eqnarray}\label{eq:unitvecdef}
\eL &=& \frac{\plep}{|\plep|}~,
\nonumber\\
\eN &=&\frac{\pq \times \plep}{|\pq \times \plep|}~,
\\
\eT &=& \eN \times \eL ~,
\nonumber
\end{eqnarray}
where $\pq$ and $\plep$ are the three momentum vectors of the
quark and the $l^-$ lepton, respectively, in the $l^+ l^-$ center-of-mass
system.
For a given lepton $l^-$ 
spin direction $\vecn$, which is a unit vector in the $l^-$ rest
frame, the differential decay spectrum is of the form
\cite{Kruger:1996cv}
\begin{equation}
\frac{d\Gamma(\shat, \vecn)}{d\shat} = \frac{1}{2}
\left(\frac{d\Gamma(\shat)}{d\shat}\right)_0 
\Big[ 1 + (\PL \eL + \PT \eT + \PN \eN) \cdot \vecn \Big],
\label{eq:poldecay}
\end{equation}
where the unpolarized decay rate 
$\left(\tfrac{d\Gamma(\shat)}{d\shat}\right)_0$
can be found in Eq.\ \eqref{diffspec}
and 
the polarisation components 
$\PI$ (${\rm i = L,N,T}$) are obtained from the relation
\begin{equation}
\PI(\shat) =  \frac{d \Gamma (\vecn = \ei)/d \shat  -  d
\Gamma (\vecn = -\ei)/d\shat} {d \Gamma (\vecn = \ei)/d\shat  
+  d \Gamma (\vecn = -\ei)/d\shat}~ .
\label{leppolldef}
\end{equation}

The resulting polarisation asymmetries are
\begin{eqnarray}
\PL(\shat)&=& \frac{a}{\Delta(\shat)} 
\Big[12\Real({\Cviieff} C_{10}^*)(1-  \shat ) 
\nonumber\\
&&+  2\Real({\Cixeff}C_{10}^*)(1 ~ + 
\shat -2 \shat^2)\Big],
\nonumber \\ 
\PT(\shat)  &=& \frac{3 \pi \mlhat}{2 \Delta (\shat) \sqrt{\shat}}
(1-\shat) \Big[2\Real({\Cviieff} C_{10}^*) 
\nonumber\\
&&
-4 \Real({\Cviieff} {\Cixeff}^*) - \frac{4}{\shat}|{\Cviieff}|^2 
\nonumber\\
&\phantom{=}&+ \Real({\Cixeff}C_{10}^*)- |{\Cixeff}|^2 \shat\Big],
\nonumber\\ 
\PN(\shat)  &=&  \frac{3 \pi \mlhat a}{2 \Delta (\shat)}
\ (1-\shat)\ \sqrt{\shat}\ \Imag({\Cixeff}^*C_{10})~,
\label{leppollasym}
\end{eqnarray}
where we differ by a factor of 2 in $\PT$ with respect to the
results obtained in \cite{Kruger:1996cv}. The above expressions for $\PI$ 
agree with \cite{dafne} for the SM case. Clearly, the polarisation asymmetries
in Eq.\ \eqref{leppollasym} have different quadratic
combinations of the Wilson couplings and any alteration in the values
of these couplings can lead to changes in the asymmetries. Hence, these
are sensitive to new physics and can also probe the relative signs of
the couplings ${\Cviieff}$, ${\Cixeff}$ and $C_{10}$. The normal polarisation
asymmetry $\PN$ is proportional to $\Imag({\Cixeff}C_{10}^*)$ and is thus
sensitive to the absorptive part of the loop contributed by the charm
quark. 
Note also that
the transverse and normal asymmetries $\PT$ and $\PN$, respectively,  are 
proportional to
$\mlhat$ and thus the effects can be significant only for the case of
tau leptons. 


\section{CP violation}\label{sec:CP}
$CP$ violation in the decay $B \to X_{s} l^{+} l^{-}$ is 
strongly suppressed in the SM following
from the unitarity of the $CKM$ matrix. 
However, in the semi-leptonic
$B$ decay, $B \to X_{d} l^{+} l^{-}$, the $CP$
violating effects can be quite sizeable. 
The $CP$ asymmetry for this decay can be observed by measuring the partial 
decay rates for the
process and its charge conjugated process \cite{Kruger:1997dt,Ali:1998sf}. 
Before turning to a derivation of CP violating asymmetries 
for the case of polarised final state leptons 
it is helpful to recall the unpolarised case.
\subsection{Unpolarized case}
In this case as $CP$ violating rate asymmetry is given by
\begin{equation}
\ACP := \frac{\Gamma_0 - \bar\Gamma_0}{\Gamma_0 + \bar\Gamma_0}~,
\end{equation}
where
\begin{eqnarray}
\Gamma_0 &\equiv& \frac{d\Gamma}{d \shat} \equiv
\frac{d\Gamma(b\to d l^+ l^-)}{d \shat}\ , \
\nonumber\\
\bar \Gamma_0 &\equiv& \frac{d\bar\Gamma}{d \shat} \equiv
\frac{d\Gamma({\bar b} \to {\bar d} l^+ l^-)}{d \shat}~.
\label{eq:gamma0}
\end{eqnarray}
The unpolarized particle decay rate 
$\Gamma_0$ can be found in Eq.\ \eqref{diffspec}. This can be rewritten 
as a product of a real-valued
function $r(\shat)$ times the 
function $\Delta(\shat)$, given in Eq.\ \eqref{eq:defFunpoll};
$\Gamma_0(\shat) = r(\shat) \ \Delta(\shat)$. Following the prescription
of \cite{Kruger:1997dt} we may write 
the modulus of the matrix elements 
for the decay and the anti-particle decay 
as 
\begin{equation}
|M| = |A + \lambda_u B|\ , \ |\bar M| = |A + \lambda_u^* B|~,
\label{eq:CPMdef2}
\end{equation}
where the CP-violating parameter
$\lu$, entering the Wilson coupling ${\Cixeff}$, has been 
defined in Eq.\ \eqref{eq:c9eff}.
Consequently, the rate for the anti-particle decay is then given by
\be
\bar \Gamma_0 = {\Gamma_0}_{|\lu \to \lu^*}
= r(\shat) \bar \Delta(\shat)~;
~\bar\Delta=\Delta_{|{\lambda_u \to \lambda_u^*}}~.
\label{apr}
\ee
 
The CP violating asymmetry following
 Eqs.\ \eqref{diffspec} and \eqref{apr} is obtained as 
\begin{equation}
\ACP(\shat) 
= \frac{-2 \Imag( \lu) \Sigma}{\Delta + 2 \Imag( \lu) \Sigma}
\simeq -2 \Imag( \lu) \frac{\Sigma(\hat s)}{\Delta (\hat s)},
\label{eq:CPdef}
\end{equation}
in agreement with the result in \cite{Kruger:1997dt}.
In Eq.\ \eqref{eq:CPdef}, 
\begin{eqnarray}
\Sigma(\shat) &=& \Imag(\xi_1^* \xi_2)[F_3(\shat) + 
6 \mlhat^2 F_1(\shat)]
\nonumber\\
&&
+6\Imag({\Cviieff} \xi_2)F_1({\shat})(1+ \frac{2 \mlhat^2}{\shat})~.
\label{eq:CPde1}
\end{eqnarray}
If we choose the lepton $l^-$ to be polarized, the above 
$CP$ asymmetry gets modified and receives a contribution from
$C_{10}$ through the interference piece with ${\Cixeff}$ 
in $|M|^2$; see Eqs.\ \eqref{eq:poldecay} and \eqref{leppollasym}. 

\subsection{Polarized case}
In this case, we define the CP violating asymmetry as
\begin{equation}
\ACP(\vecn) := \frac{\Gamma(\vecn) - \bar\Gamma(\bar \vecn = - \vecn)}
{\Gamma_0 + \bar\Gamma_0}~,
\label{eq:acpn}
\end{equation}
where
\begin{eqnarray}
\Gamma(\vecn) &\equiv& \frac{d\Gamma(\shat,\vecn)}{d \shat}
\equiv \frac{d\Gamma(b \to d l^+ l^-(\vecn))}{d \shat}\ ,
\nonumber\\
\bar\Gamma(\bar \vecn) &\equiv& \frac{d\bar\Gamma(\shat,\bar\vecn)}{d \shat}
\equiv \frac{d\Gamma(\bar b \to \bar d l^+(\bar\vecn) l^-)}{d \shat}\ ,
\end{eqnarray}
and $\Gamma_0$, $\bar\Gamma_0$ are functions as defined earlier.
In addition, $\vecn$ is the spin direction of the lepton $l^-$ in the
$b$-decay and $\bar \vecn$ is the spin direction of the $l^+$ in the
$\bar b$-decay.
For instance, assuming CP conservation, the rate for the decay
of a $B$ to a left handed electron should be the same as the rate
for the decay of a $\bar B$ to a right handed positron.
Following similar arguments as in the preceding section one obtains
the following results for the CP violating 
asymmetries for a lepton $l^-$ with 
polarisation $\vecn = \ei$ (${\rm i = L,T,N}$) \cite{babu:2003ir}:
\begin{eqnarray}
\ACP(\vecn =\pm \ei) 
&=&\frac{1}{2} 
\left[\ACP(\hat s) \pm \frac{\Delta \PI - (\Delta \PI)_{|\lu \to \lu^*}}
{\Delta(\hat s) + \bar \Delta (\hat s)} \right]
\nonumber\\
&\equiv&
\frac{1}{2} \left[ \ACP(\hat s) \pm \delta\ACP^{\rm i} (\hat s)\right].  
\label{eq:acpei}
\end{eqnarray}
One can see that the unpolarized $CP$ violating asymmetry $A_{CP}(\hat s)$
as obtained in Eq.\ \eqref{eq:CPdef} is 
altered by the polarised quantities $\delta\ACP^{\rm i}(\hat s)$ which
we obtain as
\begin{equation}
\delta\ACP^{\rm i} (\hat s)
= \frac{-2 \Imag( \lu) \delta\Sigma^{\rm i}(\hat s)}
{\Delta (\hat s) + 2 \Imag( \lu) \Sigma(\hat s)}
\simeq -2 \Imag( \lu) \frac{\delta\Sigma^{\rm i}(\hat s)}{\Delta (\hat s)}~,
\label{eq:polCPdef}
\end{equation}
with
\begin{eqnarray}
\delta\Sigma^{\rm L}(\hat s) & = &
\Imag(C_{10}\xi_2)\ (1+\shat -2 \shat^2) \ a\ ,
\nonumber\\
\delta\Sigma^{\rm T} (\hat s)& = &
\frac{3 \pi \mlhat}{2 \sqrt{\shat}} (1-\shat)
\bigg[-2 \Imag({\Cviieff} \xi_2) 
\nonumber\\
&& + \frac{1}{2} \Imag(C_{10}\xi_2)-
\shat \Imag(\xi_1^* \xi_2)\bigg]\ ,
\nonumber\\
\delta\Sigma^{\rm N} (\hat s)& = &
\frac{3 \pi \mlhat}{2 \sqrt{\shat}} (1-\shat)
\left[\frac{\shat}{2} \Real(C_{10} \xi_2)\right] a\ ,
\label{eq:sigmai}
\end{eqnarray}
where $a$ is the threshold factor defined below Eq.\ \eqref{eq:BR}. 
Interestingly, we note that the 
asymmetries $\delta\Sigma^{\rm T}(\hat s)$ and 
$\delta\Sigma^{\rm N}(\hat s)$ 
have different combinations involving the 
imaginary and real parts of $\xi_2$.
Note also that for a given polarisation there are two 
independent observables, $\ACP(\vecn =\ei)$ and $\ACP(\vecn =- \ei)$
or, alternatively, $A_{CP}(\hat s) = \ACP(\vecn =\ei)+\ACP(\vecn =- \ei)$ 
and 
$\delta\ACP^{\rm i}(\hat s) =  \ACP(\vecn =\ei)-\ACP(\vecn =- \ei)$.

\section{Numerical analysis and discussion}
\label{sec:results}
\renewcommand{\arraystretch}{1.2}

\begin{figure}[ht]
\begin{center}
\epsfig{file=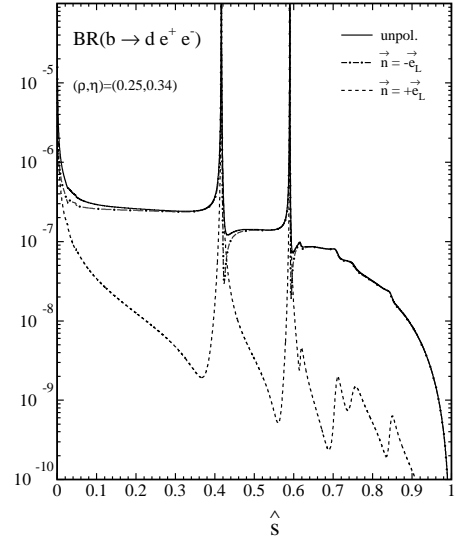,angle=0,width=7cm}
\end{center}
\caption{Polarised and unpolarised branching
ratios for the decay $b \to d e^+ e^-$
according to Eq.\ \protect\eqref{eq:BR}
and \protect\eqref{eq:poldecay}.
The unit vector $\eL$ has been defined in 
Eq.\ \protect\eqref{eq:unitvecdef}.}
\label{fig:br_e} 
\end{figure} 

\begin{figure}[ht]
\begin{center}
\epsfig{file=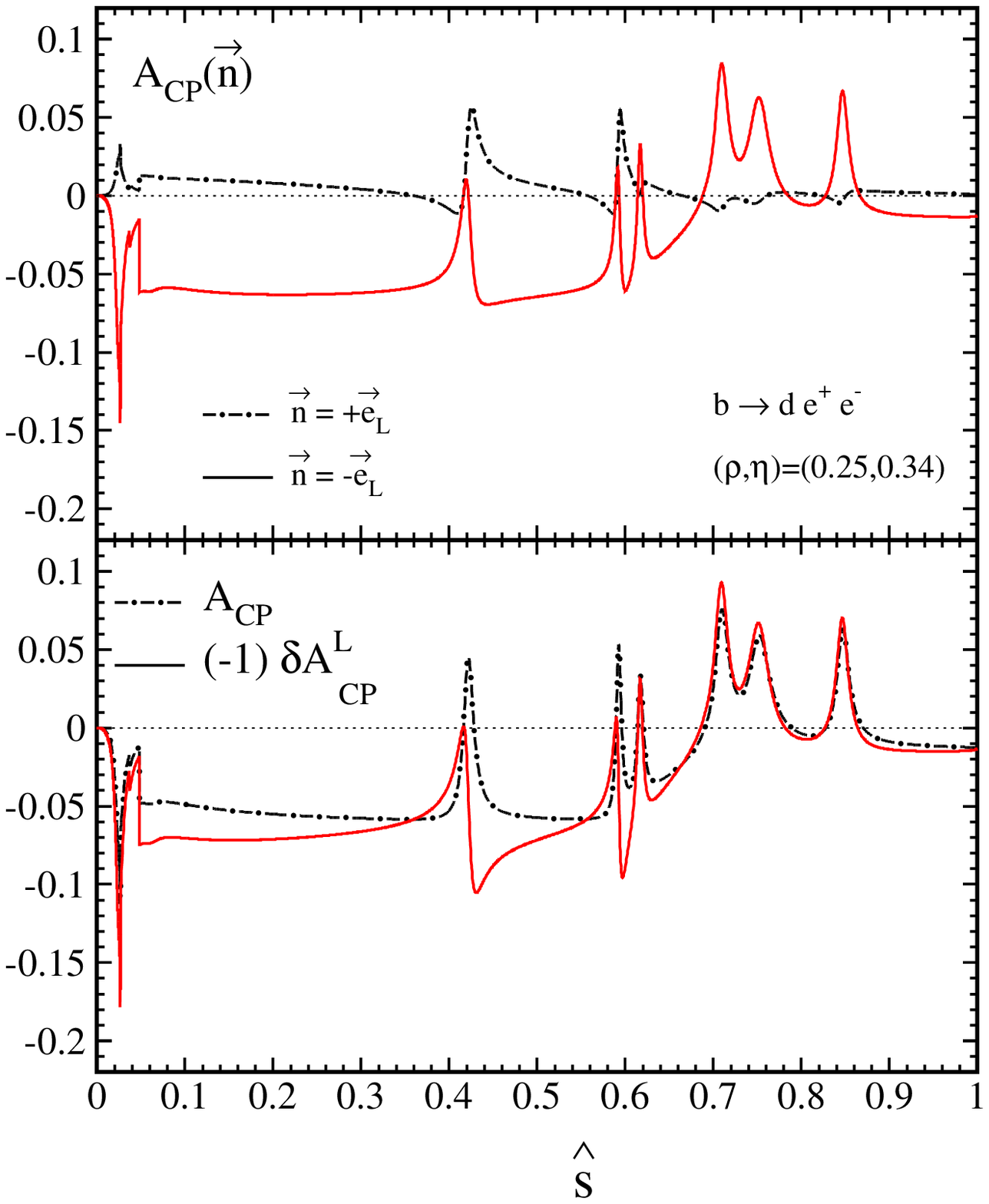,angle=0,width=7cm}
\end{center}
\caption{Polarised and unpolarised CP violating rate
asymmetries 
for the decay $b \to d e^+ e^-$
as given in Eqs.\ \protect\eqref{eq:acpei}, \protect\eqref{eq:polCPdef}
and \protect\eqref{eq:CPdef}, respectively.}
\label{fig:acp_e} 
\end{figure} 

\begin{figure}[ht]
\begin{center}
\epsfig{file=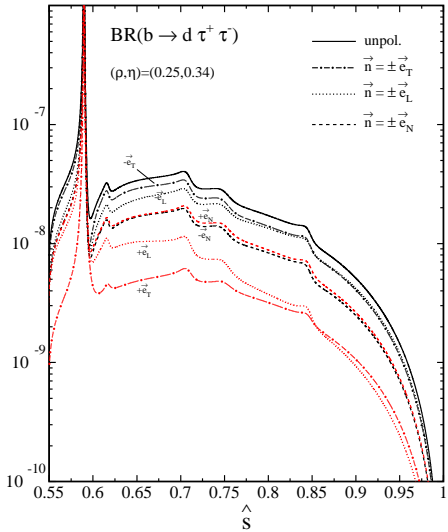,angle=0,width=7cm}
\end{center}
\caption{
As in Fig.\ \protect\ref{fig:br_e} for the
decay $b \to d \tau^+ \tau^-$.}
\label{fig:br_tau} 
\end{figure} 

\begin{figure}[ht]
\begin{center}
\hspace*{-0.5cm}
\epsfig{file=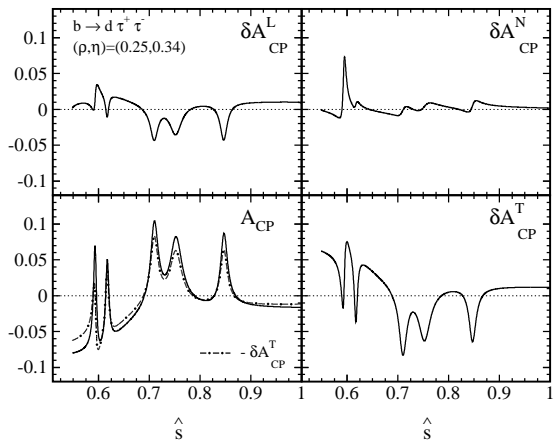,angle=0,width=8.5cm}
\end{center}
\caption{Polarised and unpolarised CP violating rate
asymmetries for the decay $b \to d \tau^+ \tau^-$
as given in Eq.\ \protect\eqref{eq:polCPdef}
and \protect\eqref{eq:CPdef}, respectively.}
\label{fig:acp_tau} 
\end{figure} 

With the above basic analytic framework, we are ready to discuss 
our numerical results. 
The basic and essential information
is summarised in figures \ref{fig:br_e} to \ref{fig:acp_tau}. 

The currently allowed range for the Wolfenstein parameters
is given by 
$0.190 < \rho < 0.268$, $0.284 < \eta < 0.366$.
For our analysis we take
$(\rho,\eta)=(0.25,0.34)$.
In terms of the Wolfenstein parameters, $\rho$ and $\eta$,
the parameter $\lu$ is given by the relation,
\be
\lu = \frac{\rho (1-\rho) - \eta^2}{(1-\rho)^2 + \eta^2}
- i \frac{\eta}{(1-\rho)^2 + \eta^2}+ \cdots \quad .
\ee
The results for other values of
these parameters can be easily obtained:
Noticing that since $\Cixeff(\shat)$ only very weakly depends on 
$\rho$ and $\eta$, almost the entire dependence is due the 
prefactors containing the CKM matrix elements; particularly,
in the expressions for the branching ratio and the CP violating 
asymmetries.
In the case of the branching ratio this is the term
$|V_{tb} V_{td}^*/V_{cb}|^2 = \lambda^2 [(1-\rho)^2 + \eta^2]$;
in the case of the CP violating asymmetries it is the factor
$\Imag \lu = -\eta/((1-\rho)^2 + \eta^2)$.
The results for other Wolfenstein parameters can therefore
be obtained by simply rescaling the shown results.
For instance varying $(\rho,\eta)$ in the allowed range
leads to a variation of $|\Imag \lu|$ in the range
$(0.54\div 0.38)$. (For $(\rho,\eta) = (0.25,0.34)$ we find
$|\Imag \lu|=0.5$.)
As can be seen the absolute value of the CP violating asymmetries 
increases with
increasing $\rho$ and $\eta$. On the other hand, the branching
ratios mildly decrease with increasing $\rho$.

 In Fig.\ \ref{fig:br_e}, we display the branching ratios for the
decay $b \to d e^+ e^-$ with unpolarized and longitudinally
polarized electrons.
The unpolarized branching ratio (solid line) has been 
obtained with help of Eq.\ \eqref{eq:BR}.
The corresponding results for polarised final state
leptons have been calculated accordingly using
Eq.\ \eqref{eq:poldecay}.
The dash-dotted and dotted lines corresponds to $\vecn = - \eL$ 
and $\vecn = + \eL$, respectively, where
the unit vector $\eL$ has been defined in Eq.\ \eqref{eq:unitvecdef}.
As can be seen, in the SM, the decay is naturally left-handed and
hence the polarised spectrum for $\vecn = -\eL$ is very similar to the
unpolarised spectrum.
In the kinematical region between the
$\rho-\omega$ and $c \bar c$ resonances, which is theoretically
the cleanest kinematic bin, the branching ratio
is $\sim 3 \times 10^{-7}$. 
On the other hand, the polarised $\vecn = \eL$ spectrum is far below the 
unpolarised one. 
This feature can provide for measurements involving
a new physics search. In the following we will classify
such polarized decays whose SM decay width is much smaller
than the unpolarized spectrum as {\it wrong sign} decay.

In Fig.\ \ref{fig:acp_e}, we present results for the polarised
and unpolarised CP violating rate asymmetries calculated according
to Eqs.\ \eqref{eq:acpei}, \eqref{eq:polCPdef}, and \eqref{eq:sigmai}
for the decay $b \to d e^+ e^-$.
We find that the asymmetries for $b \to d \mu^+ \mu^-$
are numerically similar to the results shown here, and hence are not
presented.
As mentioned earlier, only two of the shown four quantities are
linearly independent. 
As can be observed, $\ACP(\vecn = - \eL)$ is much larger than the asymmetry
with opposite lepton polarization implying 
that $\ACP(\vecn = - \eL)$ is quite similar to
the unpolarized asymmetry $\ACP$.
This can be also seen by the lower half of Fig.\ \ref{fig:acp_e} 
where the unpolarized $\ACP$ and $(-1) \ACPL$ have been plotted.
Here, $\ACP(\vecn = - \eL)$
would be the average of the two curves (lying in the middle between them). 
Note that the polarised CP violating asymmetry $\ACPL$
is comparable and in certain kinematic regions even larger than its 
unpolarized counterpart.
Particularly, in the theoretically clean region, between the 
$\rho$--$\omega$ and the $c\bar c$ resonances, we find $\ACPL$ is about 
$8 \%$ when compared to about $5 \%$ in the 
unpolarized case \cite{Kruger:1997dt,Ali:1998sf}. 
However, in the resonance regions, the polarised asymmetry can reach
values as large as up to $20 \%$ ($\rho$--$\omega$) and 
$11 \%$ ($c\bar c$), respectively.

The polarised asymmetries $\ACPN$ and $\ACPT$ are proportional to the 
lepton mass and therefore only relevant in the case of final state tau 
leptons.
In Fig.\ \ref{fig:br_tau}, branching ratios for the
decay $b \to d \tau^+ \tau^-$ are shown for unpolarized (solid line),
longitudinally (dotted), normally (dashed), and transversely 
(dash-dotted) polarized taus. The corresponding branching ratio is 
$\sim \Ord(1 \times 10^{-8})$, requiring a larger luminosity.
One can see that for $\vecn = \pm \eN$ both rates are 
very similar, whereas, for $\vecn = \pm \eT$, the $-\eT$ state is strongly 
favored, as being closer to the unpolarised decay width. Therefore, we would
classify the polarised $\vecn = \eT$ spectra as 
a {\it wrong sign} decay.

In Fig.\ \ref{fig:acp_tau}, we show both the polarised
and unpolarised CP violating rate asymmetries calculated according
to Eqs.\ \eqref{eq:acpei}, \eqref{eq:polCPdef}, and \eqref{eq:sigmai}
for the decay $b \to d \tau^+ \tau^-$. 
Since $\ACPL$ and $\ACPN$ are small we conclude that
$\ACP(\vecn = +\eL) \simeq \ACP(\vecn = -\eL)$ and
$\ACP(\vecn = +\eN) \simeq \ACP(\vecn = -\eN)$.
On the other hand, $\ACPT$ is comparable to the unpolarised
$\ACP$ as is indicated by the dash-dotted line in the $(2,1)$-panel.
This in turn implies that $\ACP(\vecn = +\eT)$ is very small. As can be
observed, all calculated asymmetries, reach at the maximum about $10 \%$. 

\section{Summary}
\label{sec:summary}
To conclude, we have performed a detailed study of the 
$CP$ asymmetry for the
process $b \to d l^+ l^-$ 
updating the unpolarised case and
including the case 
when one of the leptons is in a polarised state.
Our results indicate that when a lepton is in a certain polarised state
$(-\eL,-\eT)$, the decay rates are comparable to the unpolarised 
spectrum. For normally polarised leptons, both polarisations 
$\pm \eN$ give similar widths but lower than in the case 
of $-\eL$ and $-\eT$. The remaining polarisation states, which we had defined 
to be the {\it wrong sign} states, the decay rates and the 
corresponding asymmetries
are lower, when compared to the unpolarised SM results. 
For the kinematic regions 
which are away from resonance, the
polarised $CP$ asymmetries are larger than the unpolarised 
asymmetry. Furthermore, the resonance regions show a large
asymmetry and in all of our analysis, we have included the $\rho-\omega$ 
resonance states also. However, unfortunately, these results in the 
resonance region suffer from theoretical uncertainties. 

An observation of a  
3$\sigma$ signal for $A_{CP}(\hat s)$ requires about 
$\sim 10^{10}$ $B$ mesons \cite{Kruger:1997dt}.
For such a measurement a good $d$-quark
tagging is necessary to distinguish it from the much more copious decay
$b \to s l^+ l^-$ and hence
will be a challenging task at
future hadronic collider experiments like LHCb, BTeV, ATLAS or CMS 
\cite{Harnew:1999sq}. 
More dedicated experiments like Super-BABAR and Super-BELLE
should be able to achieve this goal.
BELLE and BABAR have already measured the rare decay $b \to s l^+ l^-$
which could be measured with great accuracy at these high luminosity
upgrades.
Given enough statistics,
and excellent kaon/pion identification, they may be able to 
measure $b \to d l^+ l^-$ or the exclusive process 
$B \to \rho l^+ l^-$.
In the polarized case, 
for measuring $\ACP(\vecn = - \eL)$ we need a similar number
of produced $B$-mesons, provided an efficient polarisation 
measurement is possible, since the branching rates are very much 
alike as discussed previously.

The polarisation observables are also interesting with
respect to new physics searches since 
they involve different quadratic combinations of the 
Wilson couplings as compared to unpolarised observables.
In this respect, given a real 
valued $C_{10}$ (for the SM), we 
note that the asymmetries $\delta\Sigma^{\rm T}$ and 
$\delta\Sigma^{\rm N}$ can be of relevance through the 
contributions arising from the 
real and imaginary parts of the function $\xi_2$ as can
be inferred from Eq.\ \eqref{eq:sigmai}.
In addition, due to the left-handed nature of the SM interactions
the electrons and muons are predominantly in the $-\eL$ state.
Hence measuring a muon in a {\it wrong sign} polarised 
state $(+\eL)$ can be very sensitive to new physics. 
Essentially, we need to probe polarised
$(+\eL)$ muons which we expect to be possible by (i) angular 
distribution of the decay products and (ii) through the life 
time of the $+\eL$ muons, which is enhanced as compared to 
the $-\eL$ state due to 
the dynamics of the SM interaction (left-handed). This is also evident by 
their smaller decay width as observed in Fig.\ \ref{fig:br_e}. 
The situation for the case of the
tau leptons is different and we observe that the $\eT$ polarised state 
can be most sensitive to new physics as can be seen
in Fig. \ref{fig:br_tau}.

\begin{acknowledgments}
I.\ S.\ is grateful to the organizers for the invitation
to the ICFP03 in Seoul and for financial support.
\end{acknowledgments}

%




\begin{thebibliography}{10}
\expandafter\ifx\csname bibnamefont\endcsname\relax
  \def\bibnamefont#1{#1}\fi
\expandafter\ifx\csname bibfnamefont\endcsname\relax
  \def\bibfnamefont#1{#1}\fi
\expandafter\ifx\csname url\endcsname\relax
  \def\url#1{\texttt{#1}}\fi
\expandafter\ifx\csname urlprefix\endcsname\relax\def\urlprefix{URL }\fi
\expandafter\ifx\csname bibinfo\endcsname\relax \def\bibinfo#1#2{#2}\fi
\expandafter\ifx\csname eprint\endcsname\relax \def\eprint#1{#1}\fi

\bibitem{Buras:2001pn}
\bibinfo{note}{{For a recent review, see A.\ J.\ Buras, hep-ph/0101336}}.

\bibitem{Glashow:1970gm}
\bibinfo{author}{\bibfnamefont{S.~L.} \bibnamefont{Glashow}},
  \bibinfo{author}{\bibfnamefont{J.}~\bibnamefont{Iliopoulos}},
  \bibnamefont{and} \bibinfo{author}{\bibfnamefont{L.}~\bibnamefont{Maiani}},
  \bibinfo{journal}{Phys. Rev.} \textbf{\bibinfo{volume}{D2}},
  \bibinfo{pages}{1285} (\bibinfo{year}{1970}).

\bibitem{Battaglia:2003in}
\bibinfo{author}{\bibfnamefont{M.}~\bibnamefont{Battaglia}} \emph{et~al.},
  \emph{\bibinfo{title}{The CKM matrix and the unitarity triangle}},
  \eprint{hep-ph/0304132}.

\bibitem{Gambino:2001ew}
\bibinfo{author}{\bibfnamefont{P.}~\bibnamefont{Gambino}} \bibnamefont{and}
  \bibinfo{author}{\bibfnamefont{M.}~\bibnamefont{Misiak}},
  \bibinfo{journal}{Nucl. Phys.} \textbf{\bibinfo{volume}{B611}},
  \bibinfo{pages}{338} (\bibinfo{year}{2001}), \eprint{hep-ph/0104034}.

\bibitem{Buras:2002tp}
\bibinfo{author}{\bibfnamefont{A.~J.} \bibnamefont{Buras}},
  \bibinfo{author}{\bibfnamefont{A.}~\bibnamefont{Czarnecki}},
  \bibinfo{author}{\bibfnamefont{M.}~\bibnamefont{Misiak}}, \bibnamefont{and}
  \bibinfo{author}{\bibfnamefont{J.}~\bibnamefont{Urban}},
  \bibinfo{journal}{Nucl. Phys.} \textbf{\bibinfo{volume}{B631}},
  \bibinfo{pages}{219} (\bibinfo{year}{2002}), \eprint{hep-ph/0203135}.

\bibitem{Glenn:1998gh}
\bibinfo{author}{\bibfnamefont{S.}~\bibnamefont{Glenn}} \emph{et~al.}
  (\bibinfo{collaboration}{CLEO}), \bibinfo{journal}{Phys. Rev. Lett.}
  \textbf{\bibinfo{volume}{80}}, \bibinfo{pages}{2289} (\bibinfo{year}{1998}),
  \eprint{hep-ex/9710003}.

\bibitem{Ali:1997bm}
\bibinfo{author}{\bibfnamefont{A.}~\bibnamefont{Ali}},
  \bibinfo{author}{\bibfnamefont{G.}~\bibnamefont{Hiller}},
  \bibinfo{author}{\bibfnamefont{L.~T.} \bibnamefont{Handoko}},
  \bibnamefont{and} \bibinfo{author}{\bibfnamefont{T.}~\bibnamefont{Morozumi}},
  \bibinfo{journal}{Phys. Rev.} \textbf{\bibinfo{volume}{D55}},
  \bibinfo{pages}{4105} (\bibinfo{year}{1997}), \eprint{hep-ph/9609449}.

\bibitem{Kaneko:2002mr}
\bibinfo{author}{\bibfnamefont{J.}~\bibnamefont{Kaneko}} \emph{et~al.}
  (\bibinfo{collaboration}{Belle}), \eprint{hep-ex/0208029}.

\bibitem{babu:2003ir}
\bibinfo{author}{\bibfnamefont{K.~S.} \bibnamefont{Babu}},
  \bibinfo{author}{\bibfnamefont{K.~R.~S.} \bibnamefont{Balaji}},
  \bibnamefont{and}
  \bibinfo{author}{\bibfnamefont{I.}~\bibnamefont{Schienbein}},
  \bibinfo{journal}{Phys. Rev.} \textbf{\bibinfo{volume}{D68}},
  \bibinfo{pages}{014021} (\bibinfo{year}{2003}), \eprint{hep-ph/0304077}.

\bibitem{Kruger:1997dt}
\bibinfo{author}{\bibfnamefont{F.}~\bibnamefont{Kr{\"u}ger}} \bibnamefont{and}
  \bibinfo{author}{\bibfnamefont{L.~M.} \bibnamefont{Sehgal}},
  \bibinfo{journal}{Phys. Rev.} \textbf{\bibinfo{volume}{D55}},
  \bibinfo{pages}{2799} (\bibinfo{year}{1997}), \eprint{hep-ph/9608361}.

\bibitem{Ali:1998sf}
\bibinfo{author}{\bibfnamefont{A.}~\bibnamefont{Ali}} \bibnamefont{and}
  \bibinfo{author}{\bibfnamefont{G.}~\bibnamefont{Hiller}},
  \bibinfo{journal}{Eur. Phys. J.} \textbf{\bibinfo{volume}{C8}},
  \bibinfo{pages}{619} (\bibinfo{year}{1999}), \eprint{hep-ph/9812267}.

\bibitem{Aliev:2003hw}
\bibinfo{author}{\bibfnamefont{T.~M.} \bibnamefont{Aliev}},
  \bibinfo{author}{\bibfnamefont{V.}~\bibnamefont{Bashiry}}, \bibnamefont{and}
  \bibinfo{author}{\bibfnamefont{M.}~\bibnamefont{Savci}},
  \eprint{hep-ph/0308069}.

\bibitem{Buras:1995dj}
\bibinfo{author}{\bibfnamefont{A.~J.} \bibnamefont{Buras}} \bibnamefont{and}
  \bibinfo{author}{\bibfnamefont{M.}~\bibnamefont{M{\"u}nz}},
  \bibinfo{journal}{Phys. Rev.} \textbf{\bibinfo{volume}{D52}},
  \bibinfo{pages}{186} (\bibinfo{year}{1995}), \eprint{hep-ph/9501281}.

\bibitem{Buchalla:1996vs}
\bibinfo{author}{\bibfnamefont{G.}~\bibnamefont{Buchalla}},
  \bibinfo{author}{\bibfnamefont{A.~J.} \bibnamefont{Buras}}, \bibnamefont{and}
  \bibinfo{author}{\bibfnamefont{M.~E.} \bibnamefont{Lautenbacher}},
  \bibinfo{journal}{Rev. Mod. Phys.} \textbf{\bibinfo{volume}{68}},
  \bibinfo{pages}{1125} (\bibinfo{year}{1996}), \eprint{hep-ph/9512380}.

\bibitem{misiak}
\bibinfo{note}{{M.\ Misiak, Nucl. Phys. {\bf B439}, 461(E) (1995)}}.

\bibitem{jezabek:1989ja}
\bibinfo{author}{\bibfnamefont{M.}~\bibnamefont{Jezabek}} \bibnamefont{and}
  \bibinfo{author}{\bibfnamefont{J.~H.} \bibnamefont{K{\"u}hn}},
  \bibinfo{journal}{Nucl. Phys.} \textbf{\bibinfo{volume}{B320}},
  \bibinfo{pages}{20} (\bibinfo{year}{1989}).

\bibitem{Falk:1994dh}
\bibinfo{author}{\bibfnamefont{A.~F.} \bibnamefont{Falk}},
  \bibinfo{author}{\bibfnamefont{M.~E.} \bibnamefont{Luke}}, \bibnamefont{and}
  \bibinfo{author}{\bibfnamefont{M.~J.} \bibnamefont{Savage}},
  \bibinfo{journal}{Phys. Rev.} \textbf{\bibinfo{volume}{D49}},
  \bibinfo{pages}{3367} (\bibinfo{year}{1994}), \eprint{hep-ph/9308288}.

\bibitem{Hewett:1996dk}
\bibinfo{author}{\bibfnamefont{J.~L.} \bibnamefont{Hewett}},
  \bibinfo{journal}{Phys. Rev.} \textbf{\bibinfo{volume}{D53}},
  \bibinfo{pages}{4964} (\bibinfo{year}{1996}), \eprint{hep-ph/9506289}.

\bibitem{Kruger:1996cv}
\bibinfo{author}{\bibfnamefont{F.}~\bibnamefont{Kr{\"u}ger}} \bibnamefont{and}
  \bibinfo{author}{\bibfnamefont{L.~M.} \bibnamefont{Sehgal}},
  \bibinfo{journal}{Phys. Lett.} \textbf{\bibinfo{volume}{B380}},
  \bibinfo{pages}{199} (\bibinfo{year}{1996}), \eprint{hep-ph/9603237}.

\bibitem{dafne}
\bibinfo{note}{{See for example, D.\ Guetta and E.\ Nardi, Phys. Rev. {\bf
  D58}, 012001 (1998)}}.

\bibitem{Harnew:1999sq}
\bibinfo{author}{\bibfnamefont{N.}~\bibnamefont{Harnew}},
  \emph{\bibinfo{title}{The B physics potential of LHCb, BTeV, ATLAS and CMS}},
  \bibinfo{note}{{Prepared for 8th International Symposium on Heavy Flavor
  Physics (Heavy Flavors 8), Southampton, England, 25-29 Jul 1999}}.

\end{thebibliography}

\end{document}